\documentclass[dvips]{acta}
\usepackage{supertabular,lscape,epsfig}
\usepackage{amssymb}
\usepackage{amsmath}
\usepackage{float}
\mathchardef\mhyphen="2D
\usepackage{placeins}%\Floatbarier
\DeclareSymbolFont{ppa}{OT1}{ppl}{m}{it}
\DeclareMathSymbol{\vv}{\mathalpha}{ppa}{'166}

\SetPages{1}{14}

\begin{document}
\newcommand\pvalue{\mathop{p\mhyphen {\rm value}}}
%Zwarte naglowki, jeden wiersz, appendix
\newcommand{\TabApp}[2]{\begin{center}\parbox[t]{#1}{\centerline{
  {\bf Appendix}}
  \vskip2mm
  \centerline{\small {\spaceskip 2pt plus 1pt minus 1pt T a b l e}
  \refstepcounter{table}\thetable}
  \vskip2mm
  \centerline{\footnotesize #2}}
  \vskip3mm
\end{center}}

%Zwarte naglowki, jeden wiersz
\newcommand{\TabCapp}[2]{\begin{center}\parbox[t]{#1}{\centerline{
  \small {\spaceskip 2pt plus 1pt minus 1pt T a b l e}
  \refstepcounter{table}\thetable}
  \vskip2mm
  \centerline{\footnotesize #2}}
  \vskip3mm
\end{center}}

%Zwarte naglowki, dwa wiersze
\newcommand{\TTabCap}[3]{\begin{center}\parbox[t]{#1}{\centerline{
  \small {\spaceskip 2pt plus 1pt minus 1pt T a b l e}
  \refstepcounter{table}\thetable}
  \vskip2mm
  \centerline{\footnotesize #2}
  \centerline{\footnotesize #3}}
  \vskip1mm
\end{center}}

%wewnątrz minipage
\newcommand{\MakeTableH}[4]{\begin{table}[H]\TabCap{#2}{#3}
  \begin{center} \TableFont \begin{tabular}{#1} #4 
  \end{tabular}\end{center}\end{table}}

%Zwarte naglowki, jeden wiersz, appendix
\newcommand{\MakeTableApp}[4]{\begin{table}[p]\TabApp{#2}{#3}
  \begin{center} \TableFont \begin{tabular}{#1} #4 
  \end{tabular}\end{center}\end{table}}

%Zwarte naglowki, jeden wiersz
\newcommand{\MakeTableSepp}[4]{\begin{table}[p]\TabCapp{#2}{#3}
  \begin{center} \TableFont \begin{tabular}{#1} #4 
  \end{tabular}\end{center}\end{table}}

%Zwarte naglowki, jeden wiersz
\newcommand{\MakeTableee}[4]{\begin{table}[htb]\TabCapp{#2}{#3}
  \begin{center} \TableFont \begin{tabular}{#1} #4
  \end{tabular}\end{center}\end{table}}

%Zwarte naglowki, dwa wiersze
\newcommand{\MakeTablee}[5]{\begin{table}[htb]\TTabCap{#2}{#3}{#4}
  \begin{center} \TableFont \begin{tabular}{#1} #5 
  \end{tabular}\end{center}\end{table}}

%Tabela w określonym miejscu
%\newcommand{\MakeTableH}[4]{\begin{table}[H]\TabCap{#2}{#3}
%  \begin{center} \TableFont \begin{tabular}{#1} #4 
%  \end{tabular}\end{center}\end{table}}

%Tabela w określonym miejscu, zwatre nagłówki, jeden wiersz
\newcommand{\MakeTableHH}[4]{\begin{table}[H]\TabCapp{#2}{#3}
  \begin{center} \TableFont \begin{tabular}{#1} #4 
  \end{tabular}\end{center}\end{table}}

%wyrównanie w tabeli - względem kropki dziesiętnej r@.l (a2_19/osb - ttt9)
%r@{\hspace{3pt}}@{$\pm$}@{\hspace{3pt}}l (a2_20/jar)
%{\it Acta Astronomica Archive}
%\parskip=0pt \itemsep=1mm \setlength{\itemsep}{0.4mm}\setlength{\parindent}{-1em} \setlength{\itemindent}{-1em} - po \begin{itemize} - wszystko
%FWHM, PSF, S/N - proste, 
%MgII, H$\alpha$
%rms, rhs, sd - kursywa
%{\sc DAOPhot}
%{\sc Fnpeaks}
%{\sf files}
%Galactic wszystko (bulge, center, plane, disk, coordinates, latitudes...)
%Cepheids
%type~ Cepheids, Population~II Cepheids
%a.u. => au (od AcA 3/2018)
%Polish National Science Centre
\newfont{\bb}{ptmbi8t at 12pt}
\newfont{\bbb}{cmbxti10}
\newfont{\bbbb}{cmbxti10 at 9pt}
\newcommand{\uprule}{\rule{0pt}{2.5ex}}
\newcommand{\douprule}{\rule[-2ex]{0pt}{4.5ex}}
\newcommand{\dorule}{\rule[-2ex]{0pt}{2ex}}
\def\thefootnote{\fnsymbol{footnote}}
\begin{Titlepage}
\Title{\bf FRBs: the Dispersion Measure of Host Galaxies}
\vspace*{5pt}
\Author{M.~~J~a~r~o~s~z~y~n~s~k~i}
{Astronomical Observatory, University of Warsaw, Al.~Ujazdowskie~4,~00-478~Warszawa, Poland}
\vspace*{5pt}
%\Received{July 17, 2020}
\end{Titlepage}

\vspace*{5pt}
\Abstract{Using the results of the {\sf IllustrisTNG} simulation we
  estimate the dispersion measure which may be attributed to halos of
  so called host galaxies of fast radio bursts sources (FRBs). Our
  results show that in contradiction to assumptions used to show the
  applicability of FRBs to cosmological tests, both the dispersion
  measure and its standard deviation calculated for host galaxies with
  given stellar mass in general increase with the redshift. The effect
  is not strong and cosmological tests using FRBs will be possible,
  but to preserve the level of statistical uncertainty the number of
  FRBs with known redshift in a sample should be increased by
  15\%--35\% depending on circumstances.  We show various statistical
  characteristics of ionized gas surrounding galaxies, the resulting
  dispersion measure and their dependence on the host galaxy stellar
  mass, redshift, and the projected distance of a FRB source from its
  host center.}{Cosmology: theory -- Galaxies: halos -- large-scale structure of Universe}

\vspace*{9pt}
\Section{Introduction}
\vspace*{5pt} Fast radio bursts (FRB) (Lorimer \etal 2007) are short events
of $\approx1$~ms duration observed at $\approx1$~GHz radio
frequencies. They are astrophysical phenomena of as yet unknown origin (see
the review by Cordes and Chaterjee 2019 -- CC19, and references
therein). The number of known FRBs exceeds one hundred\footnote{\it
  http://frbcat.org}, and it is likely to grow substantially in the near
future.

The propagation effects related to FRB allow in principle estimating
the dispersion measure (${\rm DM}=\int n_e\dd s$), the rotation measure
(${\rm RM}=\int n_eB_{||}\dd s$), and the emission measure
(${\rm EM}=\int n_e^2\dd s$) which influence the delay of the arrival time
$\tau$ of the signal at low frequencies relative to high frequencies
in proportion to $\nu^{-2}$, $\nu^{-3}$, and $\nu^{-4}$.  The
integrals are along the propagation path, $s$ measures the distance,
$n_e$ is the concentration of free electrons, and $B_{||}$ is the
magnetic field component along the ray. The term proportional to
$\nu^{-2}$ is the easiest to measure and allows finding DM
(Thornton \etal 2013). The rotation measure (RM) and the emission
measure based on optical observations may be found in some cases (\eg
in FRB 190608, Chittidi \etal 2020).

At high Galactic latitudes DM for FRBs is systematically higher than for
pulsars (Thornton \etal 2013, CC19), which suggests their extra-Galactic
origin. This, in turn, enables some cosmological tests using FRBs (\cf Gao,
Li and Zhang 2014, Zhou \etal 2014, Lorimer 2016, Yu and Wang 2017, Walters
\etal 2018). To use FRBs in cosmological tests one needs the redshifts of
the sources, which cannot be measured using radio data alone and is done by
finding a so called host galaxy of the source, where the burst was localized.

The localizations of the FRB sources, based on their positions on the sky
and proximity to possible host galaxies, are reliably known in $\approx10$
cases. Macquart \etal (2020) report localizations of four bursts discovered
by them and note another five localized by other authors -- see the
references therein. The host galaxies are of different morphological
types. The sample of five bursts used by Macquart \etal (2020) in their
analysis of the baryon content of the Universe have redshifts in the range
of 0.118--0.522. The sources lie at the projected distances of 1.5--7.0~kpc
from centers of their host galaxies.
 
The proposed cosmological tests using FRBs are mostly based on the
redshift--dispersion measure relation for the cosmological part of
the dispersion measure (${\rm DM}_{\rm cosm}(z)$). The observed ${\rm DM}_{\rm
obs}$ includes also other parts (Deng and Zhang 2014):
$${\rm DM}_{\rm obs}=DM_{\rm MW}+{\rm DM}_{\rm cosm}+{\rm DM}_{\rm
  host}/(1+z)+{\rm DM}_{\rm source}/(1+z)\eqno(1)$$
where ${\rm DM}_{\rm \rm MW}$ is the contribution from the Milky Way,
${\rm DM}_{\rm cosm}$ is due to the Inter Galactic Medium (IGM) and
possible intervening halos of galaxies close to the line-of-sight,
${\rm DM}_{\rm host}$ represents its locally measured value due to the
ionized gas in the host galaxy, and ${\rm DM}_{\rm source}$ (again in the
source frame) represents the impact of the source immediate
environment, which may contain dense plasma, as suggested by some
theoretical models (\eg Metzger \etal 2019).  ${\rm DM}_{\rm MW}$ is
usually assumed to be easy to estimate based on the map of Galactic
pulsars dispersion measure, but may also contain some unknown part
related to the Galaxy halo (Prochaska and Zheng 2019). ${\rm DM}_{\rm
source}$ can be estimated based on the optical observations of the
source environment and rotation measure of the burst as in the case of
FRB 190608 (Chittidi \etal 2020). ${\rm DM}_{\rm host}$ (which excludes
the source vicinity) can be estimated only statistically, based on the
host properties and the location of the source relative to the host
center.

In a conservative approach to the cosmological tests based on FRBs (\eg
Yang and Zhang 2016, Walters \etal 2018, Jaroszynski 2019, hereafter J19)
it is assumed that the host contribution to the dispersion measure has a
normal distribution which does not depend on time with the expected value
$\langle {\rm DM}_{\rm host}\rangle\approx 200$~pc/cm$^3$ and the standard
deviation $\sigma_{\rm host}\approx50$~pc/cm$^3$ (other values were also
considered). Since the expected value of the cosmological part increases
with the redshift (approximately as ${\rm DM}_{\rm cosm}\propto z$ -- \cf
Zhang 2018) and the contributions from hosts are divided by the $1+z$
factor, their importance diminishes for far away sources, which makes
cosmological tests plausible with a relatively small samples of FRBs with
measured redshifts. Different properties of ${\rm DM}_{\rm host}$ may
require much bigger samples of the bursts for the tests.

In this article we investigate the contribution of the halos of
galaxies to the observed dispersion measure using the results of {\sf
  IllustrisTNG-100} simulation (Nelson \etal 2018). In the next
section we describe our method of finding the angle averaged
distribution of free electrons as a function of the distance from the
center of a galaxy. In Section~3 we calculate the dispersion measure
for hosts of various stellar masses at the redshifts $0\le z\le3$. We
give the results averaged over all possible locations of a source
inside a galaxy and also for sources at a given projected distance
from the galaxy center. The discussion follows in the last section.

\Section{Distribution of Free Electrons around Galaxies}
The {\sf IllustrisTNG-100} simulation (Nelson \etal 2018, Pillepich
\etal 2018, Springel \etal 2018, Naiman \etal 2018, Marinaccii \etal
2018) gives the distribution of gas and gravitationally bound halos in
a simulation cube of the size of $75/h$ Mpc. Positions of the halos,
dark matter particles, and gas cells, their masses, and other relevant
parameters are accessible from {\sf IllustrisTNG} database (Nelson
\etal 2019).  Some of the dense gas cells may become wind particles
and possibly stars according to {\sf Illustris} naming
convention. They are treated separately and constitute a dedicated
part of the database.  If the star formation has already taken place,
star particles have assigned luminosities in several filters. This
gives the opportunity to distinguish galaxies from dark halos: we
treat an object as a galaxy if it consists of a halo with star
particles within 30 kpc from its center. The diffuse gas is present in
the vicinities of so defined galaxies and around dark halos as
well. Assuming that FRBs are somehow related to stars and products of
their evolution, we limit ourselves to investigate galaxies.

Because of the technical limitations (data volume) we are able to use
the lowest resolution version of the full cube simulation named {\sf
  TNG100-3}. (Compare J19, based on {\sf Illustris-3} data for the
same reason).  Some numerical experiments with {\sf TNG100-3} show
that the number of star cells within $\approx10$~kpc around galaxy
centers is too small to give realistic maps of their distribution
there, even if we stack distributions around many galaxies. For this
reason we also use data from the simulations limited to smaller
volumes but having better spatial resolution named {\sf
  TNG100-1-Subbox0} and {\sf TNG100-1-Subbox1}. We consider several of
the snapshots corresponding to the redshifts $z=0$, 0.2, 0.5, 0.7,
1.0, 2.0, and 3.0 which should give a representative description of
changes in gas and stars distribution with time.

\newpage
For the subboxes the information on the gravitationally bound halos is
not provided. We construct low resolution ($1024^3$) 3D maps of total
matter density within subboxes and look for its local maxima. For each
maximum we investigate radial density distribution in its vicinity by
counting mass in thin spherical shells. We proceed until the averaged
density within the sphere falls to $\approx200 \rho_c$, $\rho_c$ being
the critical matter density calculated for the given epoch.  Next we
find the sphere containing half of the total mass. For the matter
inside this sphere we calculate its center of mass and repeat the
calculation of radial mass distribution around it. After a few
iterations we obtain a galaxy candidate with a given center of mass
position. It may happen that our algorithm, starting from different
local mass density maxima produces at the end more than one galaxy
candidate at the same or very close positions. We remove these extra
candidates if necessary. We also remove candidates which have no stars
and constitute dark halos. When looking for the candidate galaxies we
establish characteristics of their different component distributions:
$r_*$ -- half star radius (of a sphere containing half of the star
mass), and similarly $r_{\rm gas}$ for gas, and $r_{\rm dark}$ for
dark particles. (Same parameters are provided for the halos in {\sf
IllustrisTNG-100-3} database.)

Our goal is to estimate the distribution of the dispersion measure
related to the host galaxy for a source randomly located inside it. In
a brute force approach one would consider many source positions inside
each galaxy and many paths going from a source to infinity checking
the gas cells on the way and including their individual contributions
to the dispersion measure. Such an approach is technically challenging
(what is the contribution to the dispersion measure of a given gas
cell if a ray passes at a given distance from its center?) and not
necessary to obtain results averaged over many source positions and many
ray directions. We assume that the location of sources follows the
distribution of stars. The direction of the line-of-sight relative to
the host galaxy for a given source location depends on the observer
position.  Isotropy of the Universe implies isotropic distribution of
observers relative to the source and isotropic distribution of their
lines-of-sight. Going a step further we also assume that (if averaged
over many host galaxies) the distribution of stars and gas is
spherically symmetric relative to the galaxy mass center. Under such
simplifying assumptions it is only necessary to consider sources at
different distances from the host center and rays making different
angles with the direction toward the center, so the parameter space
which should be investigated is two-dimensional.

We expect some dependence of ${\rm DM}$ on the host properties. Such
relation should include parameters which can be measured to be useful. We
have considered the total mass of stars and luminosities in different
filters, which can be estimated with photometric observations. Since {\sf
  Illustris} database provides masses of the stellar cells and their
absolute luminosities in several filters, the synthetic galaxy parameters
can be calculated. We found the total stellar mass to be the most
convenient parameter.  We divide all galaxies into seven bins with equal
logarithmic widths (0.5~dex) covering the range $7.5\le\lg M_*\le11.0$
where $M_*$ is the total stellar mass in solar units. We also put an upper
limit on the total galaxy mass $M_{\rm tot}\le10^{13}~\MS$ to exclude
galaxy clusters and groups, which are systems of different kind.

We construct gas distributions stacking galaxies belonging to
different mass ranges separately.  We are interested in electron
density $n_e$ as a function of the distance from the galaxy center
$r$. Our grid covers the radii from 1 kpc to 10 Mpc with equal
logarithmic bins (0.1~dex). For each galaxy we consider gas cells
within $2r_{\rm gas}$. The number of free electrons in a gas cell is
given by parameters from the database:
$$N_e =A\frac{XM_{\rm gas}}{m_H}\eqno(2)$$
where $A$ is the ratio of the electrons number to the hydrogen atoms
number inside the cell, $M_{\rm gas}$ is the total gas mass, and $m_H$
is the hydrogen atom mass. In calculations we use the hydrogen mass
fraction $X=0.75$. The volume of the gas cell $V$ is also given in the
database, so the mean electron concentration $n_e=N_e/V$ can be
calculated. For further use we divide the interesting range of the
comoving electron densities (from $10^{-10}$ to $10^0$ cm$^{-3}$) into
a hundred equal logarithmic bins.  Electron concentration of each cell
belongs to one of these bins.

\begin{figure}[htb]
\vglue-3mm
\centerline{\includegraphics[width=10.4cm]{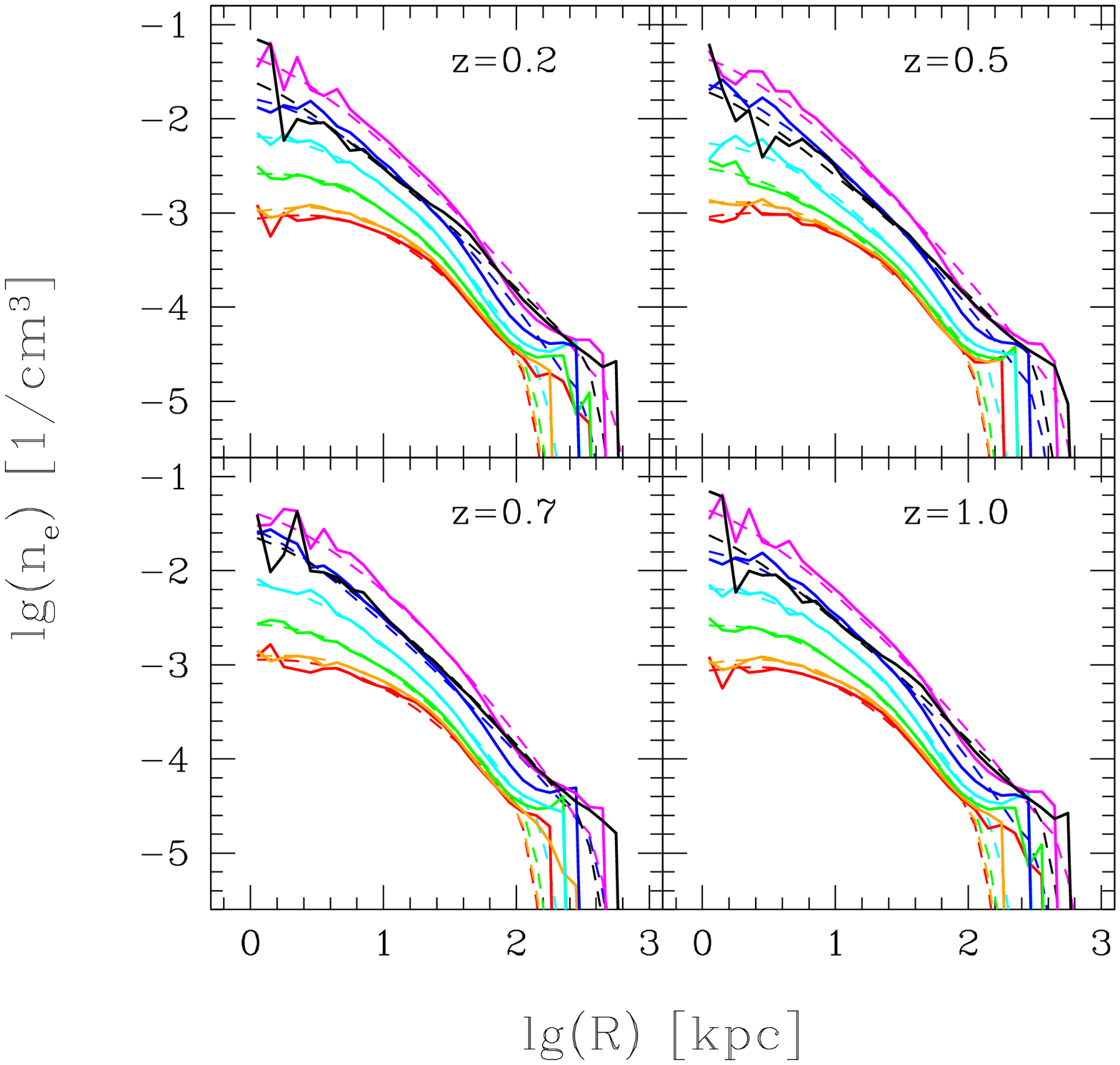}}
\vspace*{-4pt}
\FigCap{Electron concentration profiles in galaxies belonging to different 
stellar mass ranges at different redshifts (solid lines). 
The redshifts are given by the
labels and the mass ranges by colors: $\lg(M_*/\MS)\in[7.5,8.0]$ (red),
$[8.0,8.5]$ (orange),
$[8.5,9.0]$ (green),
$[9.0,9.5]$ (cyan), 
$[9.5,10.0]$ (blue), 
$[10.0,10.5]$ (magenta), and
$[10.5,11.0]$ (black).
The plots for bins 1--6 are ordered and drawn with colors. 
The plots for the highest mass bin 
cross the other and we use black color to draw them.
The rough analytical approximations to the density profiles are marked with 
dashed lines. (See text for details.) 
We use comoving distances and comoving densities; the true 
distances are smaller by a $(1+z)$ factor, and the densities are 
$(1+z)^3$ times higher.}
\end{figure}
At the starting point of the simulation gas cells have typical
comoving sizes of 75\,000/1820$h^{-1} \approx$ 41$h^{-1}$ kpc, much more
than the thickness of the spherical shells in our radial grid close to
the center. Gas cells which happen to be close to the center of a
galaxy are much smaller and denser than the average but some of them
would not fit into a single shell. We assign a gas cell to a radial
bin if its center coordinates fall within. Redistribution of gas into
adjacent radial bins is not possible since the shape of the cells is
not given. (Assuming spherical shape leads to rather unpleasant
calculations which are not worth the effort.) Instead we consider the
averaged electron density in all cells whose centers fall into a given
radial bin around a galaxy belonging to a given mass bin. We construct
a histogram ${\rm Vol}(r^i,M_*^j,n_e^k)$.  If the center of a gas cell of
volume $V$ lies in the $i$-th radial bin surrounding a galaxy
belonging to the $j$-th mass bin and has electron concentration
belonging to the $k$-th density bin, the histogram changes according
to the rule:
$${\rm Vol}(r^i,M_*^j,n_e^k) \rightarrow {\rm Vol}(r^i,M_*^j,n_e^k)+V.\eqno(3)$$
After checking positions of all gas cells relative to all galaxy centers
we calculate the averaged electron densities:
$$\left\langle n_e(r^i,M_*^j)\right\rangle=\frac{\sum_k~n_e^k{\rm Vol}(r^i,M_*^j,n_e^k)}{\sum_k~{\rm Vol}(r^i,M_*^j,n_e^k)}.\eqno(4)$$
Since all gas cells fill the whole simulation cube to a good
approximation, the density defined above is realistic: there is no empty
space around the cells which would increase the denominator and the cells 
do not overlap.

The electron density profiles for several redshifts and galaxy mass
bins are shown in Fig.~1. Due to the binning of the data and
relatively small number of gas cells close to the center, the result
is not a smooth function of $r$. It represents a typical dependence on
the radius at a discrete set of points for a halo belonging to one of
the considered mass ranges. The electron density profiles for
$M_*\le3\times10^{10}~\MS$ are ordered, \ie the density increases with
mass for a fixed radius. The highest mass range is exceptional: its
density plot crosses the plots for the lower mass ranges. At $z=3$
the effect is absent. Relatively lower density of the electrons close
to the centers of high mass galaxies implies their lower dispersion
measure (see Section~3).

For further application we find a rough analytical approximation to
the electron concentration dependence on radius.  We look for a fit of
the form:
\vspace*{-4pt}
$$y(x)= a_0+a_1x+a_2x^2\qquad\text{where}\quad
x\equiv\lg r\quad\text{and}\quad y\equiv\lg n_e\eqno(5)$$
where $a_0$, $a_1$, and $a_2$ are the fit parameters different for
each mass bin.

In our approach only gas cells within $2r_{\rm gas}$ from a galaxy center
are considered as belonging to it. Even within the single mass bin this
parameter has some scatter. On the other hand the fiducial mass density
$200\rho_{\rm crit}$ calculated for the given epoch, may serve as a typical
density at the halo boundary. At the present epoch it corresponds to the
electron concentration $n_{\rm e0}\approx4\times10^{-5}$~cm$^{-3}$ (for
baryon density parameter $\Omega_B=0.05$ and $h=0.7$).  Inspecting the
plots in Fig.~1 one can see that after reaching this density the
concentration rapidly falls off. For $x\ge x_{\rm 4.4}$, where $y(x_{\rm
  4.4})=-4.4=\lg(4\times 10^{-5})$, we use an {\it ad hoc} approximation
$y(x)=-4.4-20(x-x_{\rm 4.4})^2$ which represents a cut-off. The original
and approximated curves can be compared in Fig.~1.

\Section{Estimates of the Host Galaxy Contribution to the Dispersion
Measure}
Using the analytical approximation to the electron concentration
profile $n_e(r)$ as defined in Eqs.~4 and 5 for a given galaxy mass
range one can calculate the dispersion measure along a
line-of-sight. We define:
$${\rm DM}(R,s_0)=\int_{\rm s_0}^\infty n_e(\sqrt{R^2+s^2})\dd s\eqno(6)$$
where the line-of-sight is passing at the distance $R$ from the galaxy
center.  The length along the line-of-sight is measured by $s$ with
$s=0$ at the closest point to the center. $s_0$ defines the source
position and the integration formally goes to infinity, but in
practice electron density falls to zero at some radius (\cf
Fig.~1) characteristic for each mass range.
 
For a source at the distance $r$ from the halo center and a line-of-sight 
at the angle $\theta$ relative to the direction toward the center one has:
$${\rm DM}_{\rm host}(r,\theta)={\rm DM}(r\sin\theta,-r\cos\theta).\eqno(7)$$
Perhaps some selection mechanism makes the sources in the far away or
close part of the host more/less likely to be observed but having no
data we assume that they are distributed with spherical symmetry,
which implies uniform distribution of $\mu=\cos\theta$ in the range
$(-1,+1)$. The radial distribution of sources is also unknown. We use
a very simplified approach assuming that the sources follow the mass
distribution of stars.

Stellar cells have densities $\approx10^6$ times higher than average so
there is no problem with their excessive sizes as compared to the
thickness of radial bins in our grid. We construct histograms giving
the distribution of stellar mass inside galaxies belonging to given
mass bin, ${\rm Mass}(r^i,M_*^j)$ and the number of galaxies in each mass
bin ${\rm Num}(M_*^j)\!$. For a stellar cell of mass $M$ in a distance
belonging to the $i$-th radial bin surrounding a galaxy belonging to
the $j$-th mass bin the histograms change according to:
\setcounter{equation}{7}
\begin{eqnarray}%8-9
{\rm Mass}(r^i,M_*^j) &\rightarrow&  {\rm Mass}(r^i,M_*^j) + M \\
{\rm Num}(M_*^j)      &\rightarrow&  {\rm Num}(M_*^j) + 1 
\end{eqnarray}
After checking positions of all stellar cells relative to all galaxy
centers we obtain the averaged stellar mass in a radial bin surrounding a
galaxy belonging to given mass bin:
$$M^j(r^i)=\frac{{\rm Mass}(r^i,M_*^j)}{{\rm Num}(M_*^j)}.\eqno(10)$$
\begin{figure}
\centerline{\includegraphics[width=5.5cm]{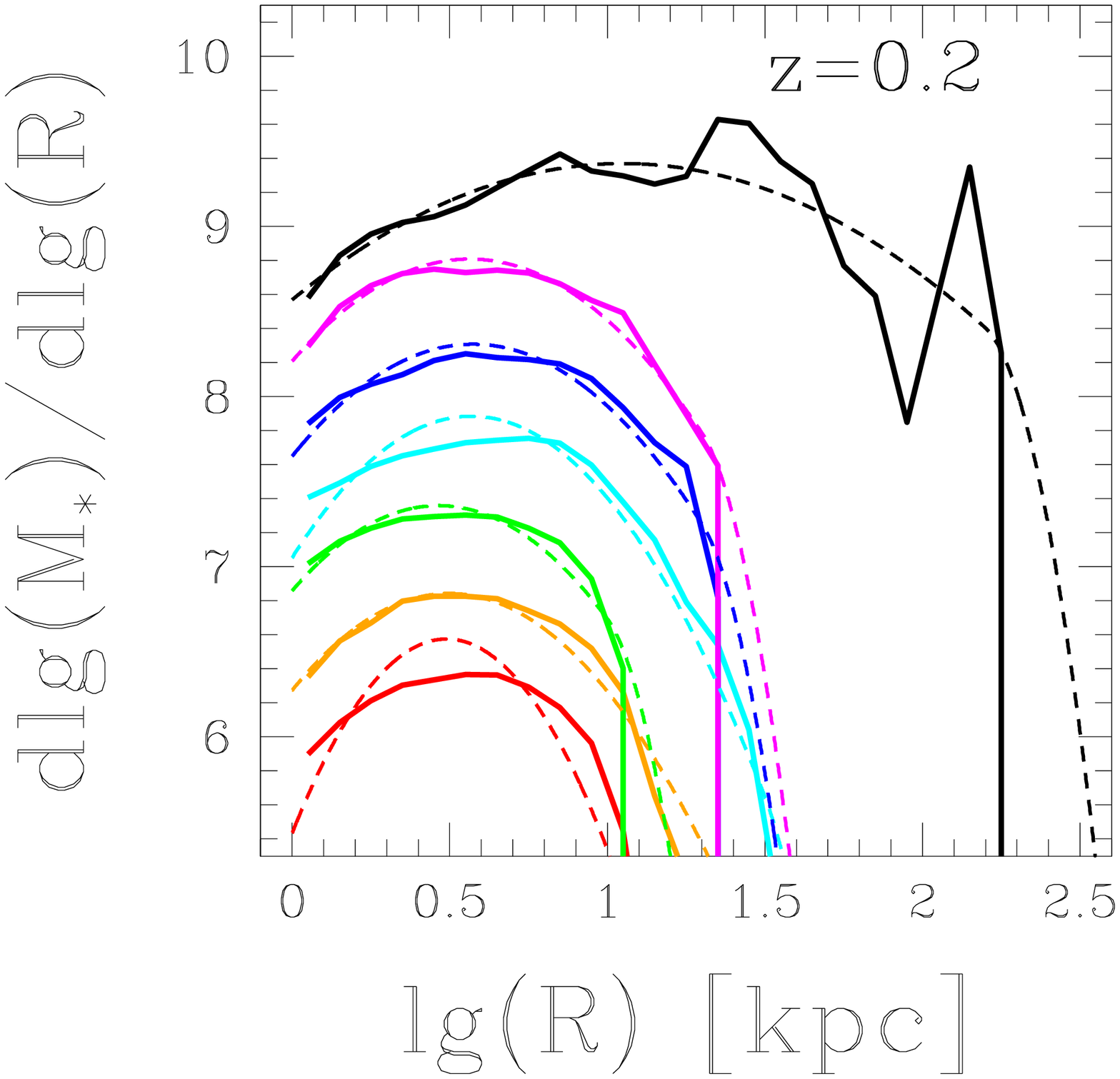}\hfill
\includegraphics[width=5.5cm]{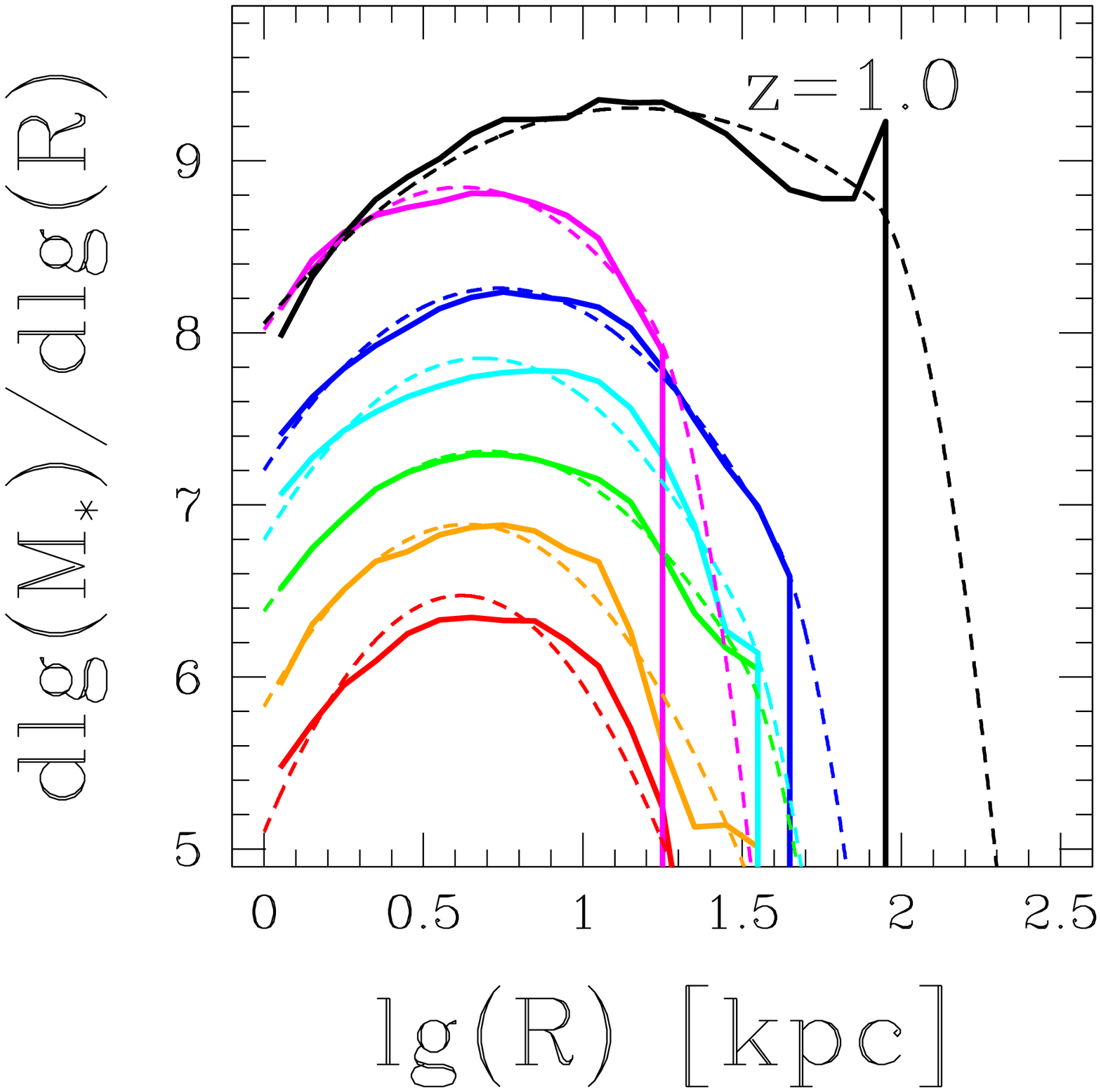}}
\FigCap{Examples of the stellar mass distribution in galaxies belonging 
to different mass ranges for $z=0.2$ and $z=1.0$ as labeled. 
Color conventions follow Fig.~1. The raw data are plotted with solid lines 
and the analytical approximation uses dashed lines. See text for details.}
\end{figure}
In our approach the mass of stars in a spherical shell around a galaxy
center is a relevant variable giving the relative probability of
finding a source within the same radius range. In Fig.~2 we show the
distribution of star mass in spherical shells of the same logarithmic
thickness (0.1~dex) around galaxies of different mass ranges based on
{\sf IllustrisTNG} results for chosen epochs. The volumes of
considered shells $V(r)\propto r^3$, so the mass distributions have
maxima at a few kiloparsecs from the center, despite the fact that the
averaged density of stars is a monotonically decreasing function of
radius.  Our analytic approximations (see Fig.~2) are obtained with a
method very similar to the method applied to the electron density
profiles.  Using the approximated radial star mass distribution
$M^j(r)$ we have the approximated star density:
$$\rho_*^j(r)=\frac{\dd M^j/\dd r}{4\pi r^2}.\eqno(11)$$

\begin{figure}[htb]
\centerline{\includegraphics[width=10.5cm]{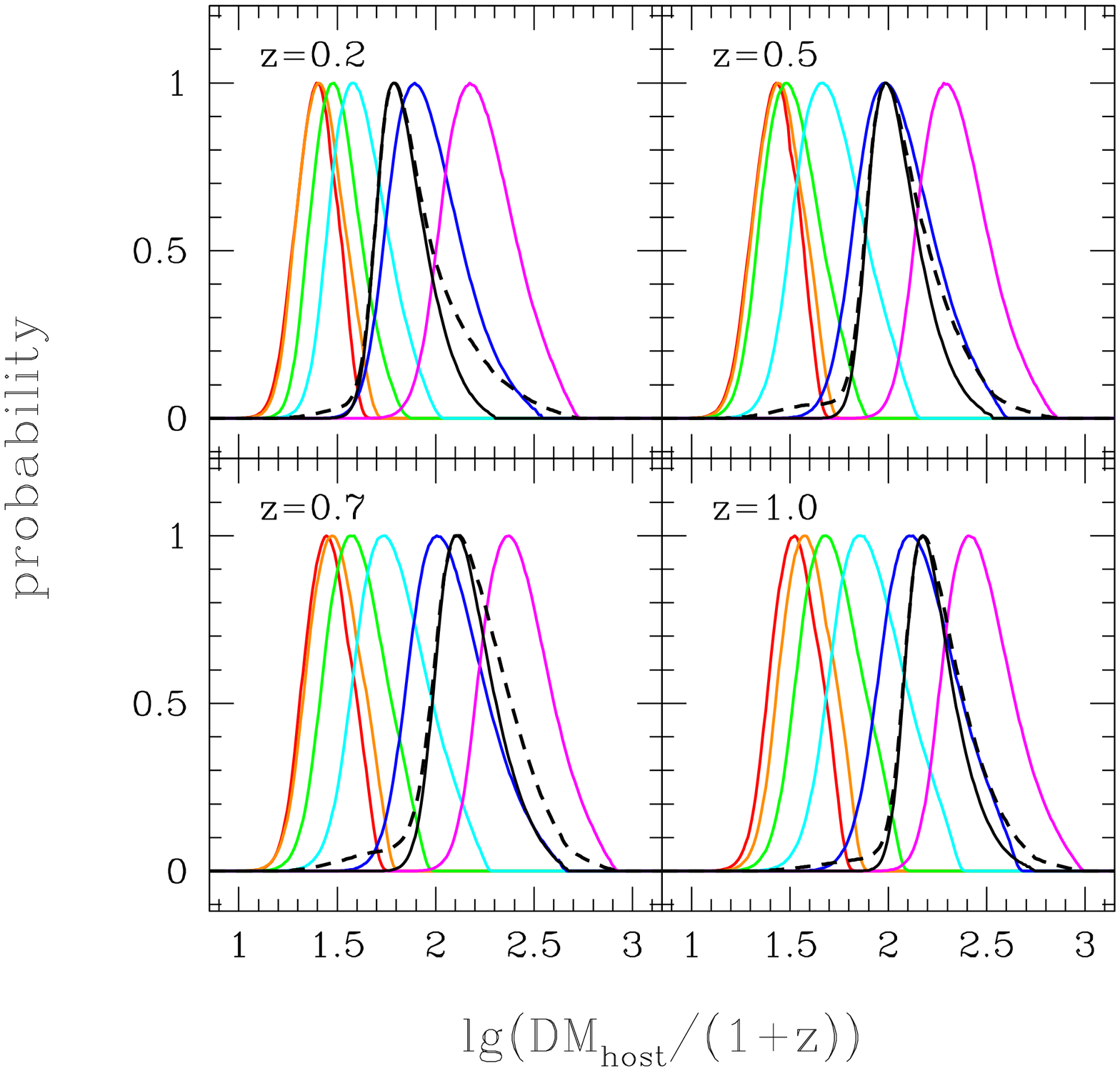}}
\FigCap{Distributions of probability for the host galaxy contribution
to the dispersion measure (DM), normalized to one at maximum. 
The color convention follows previous figures. The distribution averaged over 
seven mass bins is plotted as black dashed line. Panels correspond to different
redshifts as denoted by labels. DM-s are calculated in the rest frames of
the sources and divided by the $1+z$ factor to represent their
contribution to the total observed value. }
\end{figure}

For each galaxy mass range we calculate DM on a grid of source
positions distributed uniformly in $\lg r$ and $\mu$. Under our
assumption the probability $p(\lg r,\mu)$ of finding a source position
at any point of the grid scales as $p(\lg r,\mu)\propto \dd M_*/\dd\lg r$.
After calculating DM along many paths we obtain its probability
distributions separately for each galaxy mass range. These are
distributions resulting from averaging over all possible source
positions. They are shown in Fig.~3. To make the plots readable we
transform the density probability functions $f^j(DM)$ to have
$\max(f^j({\rm DM}))=1$, where the subscript $j$ enumerates the mass bins.
The averaged ${\rm DM}^j$ for given mass bin and its variance are given by
standard expressions:
\setcounter{equation}{11}
\begin{eqnarray}%12-14
\left<{\rm DM}^j\right>&=&\frac{1}{A^j}\int_0^\infty~{\rm DM}~f^j({\rm DM})\dd\,{\rm DM}\\
\sigma_{\rm DM}^2&=&\frac{1}{A^j}\int_0^\infty~\left({\rm DM}-\left\langle {\rm DM}^j\right\rangle\right)^2~f^j({\rm DM})\dd\,{\rm DM}\\
A^j&=&\int~f^j({\rm DM})\dd\,{\rm DM}
\end{eqnarray}
where $A^j$ is a normalizing factor. Finally we obtain the probability
distribution for all the combined mass ranges. Denoting by $M_*^j$ the
total mass of stars in galaxies belonging to the $j$-th mass bin and
by $M_*^{\rm tot}$ their sum, we have:
$$f({\rm DM})=\sum_{\rm j=1}^7~\frac{M_*^j}{M_*^{\rm tot}}\frac{1}{A^j}f^j({\rm DM})\eqno(15)$$

\begin{figure}[htb]
  \centerline{\includegraphics[width=10.5cm]{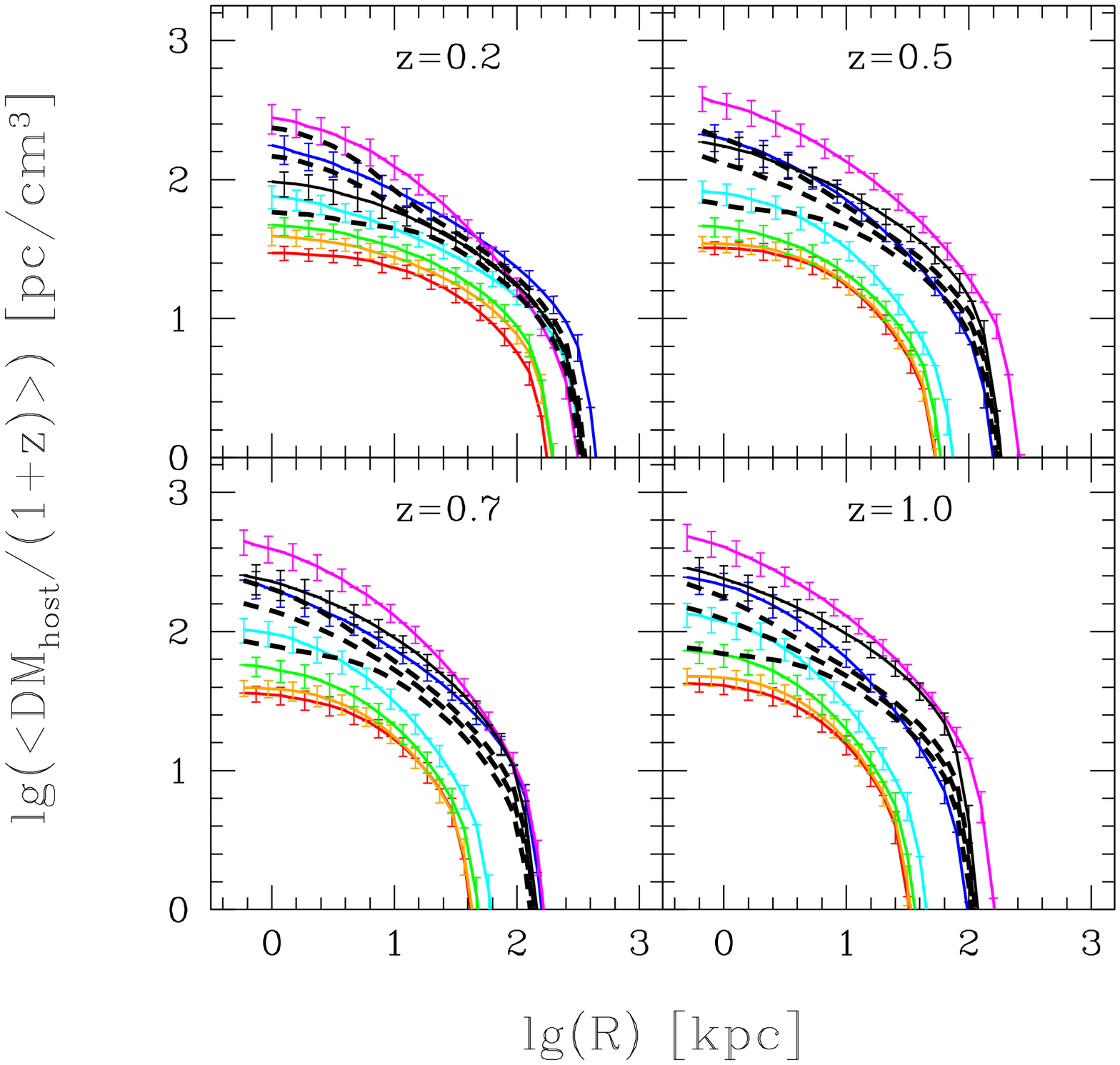}} 
  \FigCap{The host galaxy contribution to the observed dispersion
    measure as a function of its physical distance from the
    line-of-sight. Color lines with error bars give the results for
    different galaxy mass ranges averaged over source positions along
    the line-of-sight. The color convention follows previous
    figures. Results for hosts at different redshifts are shown in
    separate panels, as labeled. The averages over all mass ranges and
    their $\pm$ sigma deviations are shown using thick black dashed
    lines.}
\end{figure}
In some cases the distance between the FRB's line-of-sight and the
host center may be measured. This gives an extra constraint on the
source position inside the host galaxy. Since we assume that the
sources follow the distribution of stars and the density of stars
decreases sharply with increasing distance from the host center, the
vicinity of the point closest to the center on the line-of-sight is
the most probable position of the source. For a line-of-sight passing
at the distance $R$ from the host center and the source at position
$s_0$ along the line-of-sight the dispersion measure ${\rm DM}(R,s_0)$ is
given by Eq.(6). The probability of the source position $s_0$ at given
$R$ is proportional to the star density
$\rho_*^j\left(\sqrt{R^2+s_0^2}\right)$, so the average dispersion
measure and its variance for the $j$-th galaxy mass range are given
by:
\setcounter{equation}{15}
\begin{eqnarray}%16-18
\left\langle{\rm DM}^j(R)\right> &=& \frac{1}{B^j}\int_{-\infty}^{+\infty} {\rm DM}^j(R,s_0)\rho_*^j\left(\sqrt{R^2+s_0^2}\right)\dd s_0\\
\sigma^2_{\rm DM}           &=&\frac{1}{B^j}\int_{-\infty}^{+\infty}\left({\rm DM}^j(R,s_0)-\left\langle{\rm DM}^j(R)\right\rangle\right)^2\rho_*^j\left(\sqrt{R^2+s_0^2}\right)\dd s_0\\
B^j                         &=&{\rm Num}(M_*^j)\int_{-\infty}^{+\infty}\rho_*^j\left(\sqrt{R^2+s_0^2}\right)\dd s_0
\end{eqnarray}
where $B^j$ is a normalizing factor. The averaged over all mass ranges 
$\langle{\rm DM}(R)\rangle$ can be obtained as weighted mean of
$\langle{\rm DM}^j(R)\rangle$ with weights proportional to $B^j$.

\begin{figure}[htb]
\centerline{\includegraphics[width=5.5cm]{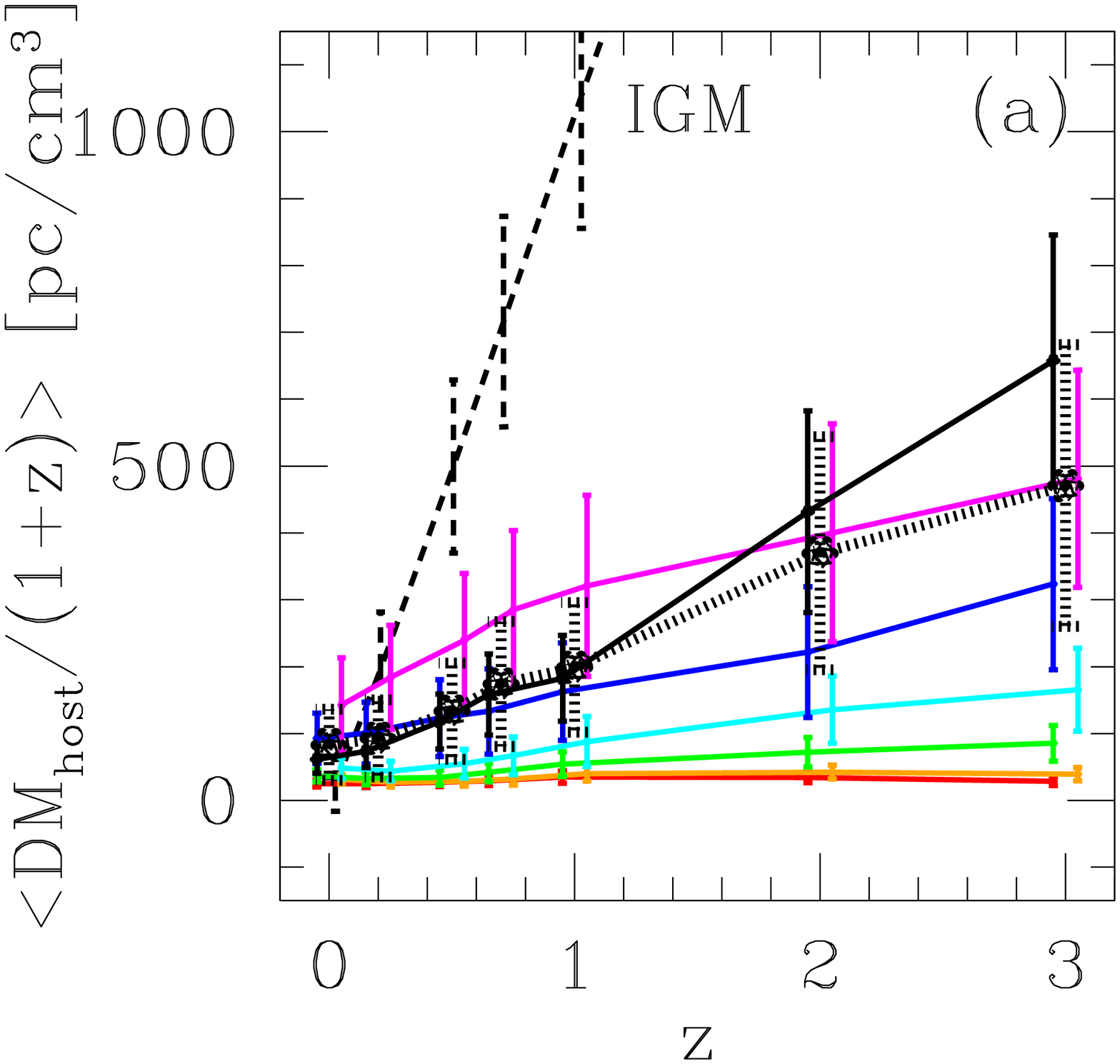} 
\hfill \includegraphics[width=5.5cm]{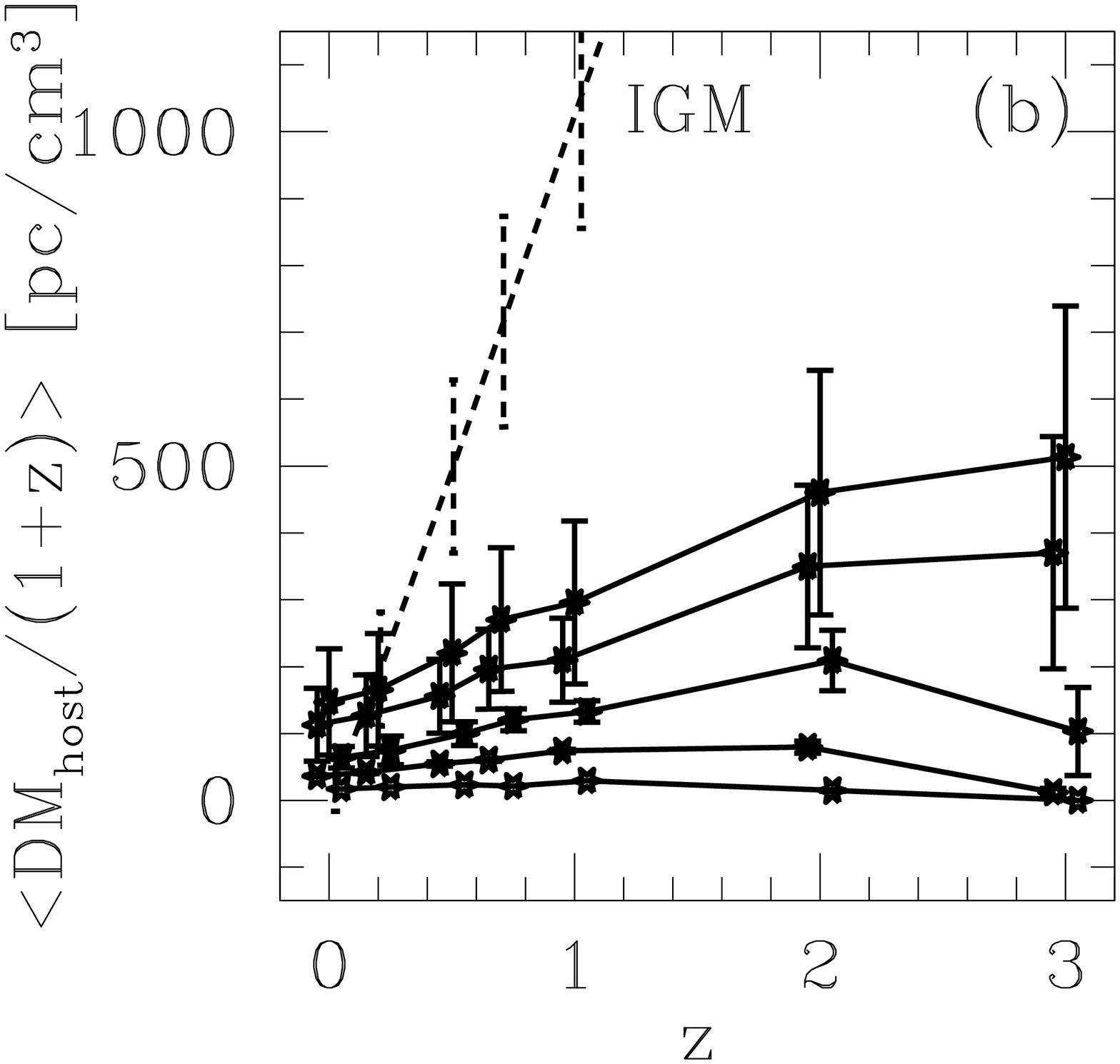}}
\FigCap{({\it a})~The source position averaged contribution of
  the host galaxy to the observed dispersion measure ${\rm DM}_{\rm
    host}/(1+z)$ and its standard deviation $\sigma_{\rm host}/(1+z)$
  (error bars) as a function of the redshift for different mass ranges
  (color conventions follow Fig.~1). The thick solid line with
  asterisks gives the dependence averaged over the whole host mass
  range.  ({\it b})~The host contribution to the observed dispersion
  measure averaged over all mass ranges and source position along the
  line-of-sight at given physical distance from the galaxy center:
  0~kpc, 3~kpc, 10~kpc, 30~kpc, and 100~kpc (top to bottom) with
  standard deviation (error bars) as a function of the redshift. In {\it
    both panels} the positions of the error bars are slightly shifted
  in redshift to avoid their overlapping. The dashed lines show the
  cumulative dispersion measure and its standard deviation due to the
  IGM between the observer and given redshift, based on the data from
  J19.}
\end{figure}
The averaged dispersion measures for all host mass bins as a function of
the projected distance from the host center $R$ are given in Fig.~4 with
standard deviations shown as error bars. In Fig.\,5a we show dispersion
measures ${\rm DM}$ and their standard deviations $\sigma_{\rm DM}$ as a
function of the redshift for different galaxy mass ranges, averaged over
all possible source positions. In Fig.\,5b we show ${\rm DM}$ and its
standard deviation as a function of the redshift for sources at known
projected distances from the galaxy center, averaged over all galaxy mass
ranges.

We also present our numerical results in Table~1. The table gives the
contributions to the observed dispersion measure as a function of the
redshift (1-st column). For comparison in the 2-nd column we give the
expected contribution from IGM, as calculated by J19. The 3-rd column
($\langle {\rm DM}\rangle$) gives the expected contribution averaged over
galaxies of all mass ranges and all possible source positions assumed to
follow the distribution of stars. The remaining five columns give the
expected contribution ${\rm DM}_R$ for a source at projected distance $R$
(where $R=0$~kpc, 3~kpc, 10~kpc, 30~kpc, and 100~kpc respectively) from a
galaxy center, averaged over all galaxy mass ranges.

Our calculations show that in general the contribution of the hosts to the
observed dispersion measure increases with the redshift despite the $1/(1+z)$
reducing factor. This can be seen in Fig.\,5a for galaxy mass ranges with
$M_*\ge 3\times10^8~\MS$.  It is also true about the standard deviations,
which may reach values of 100--200~pc/cm$^3$ at source redshifts $1\le z
\le3$. On the other hand for galaxies with small masses $M_*\le3\times
10^8~\MS$ (two low mass ranges) ${\rm DM}$ and $\sigma_{\rm DM}$ remain low.

Examining column ${\rm DM}_0$ in Table~1 one can see large values of DM 
and $\sigma_{\rm DM}$, but this is an extreme case with a source in 
projection at the very center of the host, which has a low probability. 
At $R\ge30$~kpc the contribution from the hosts becomes negligible 
as compared to the expected IGM values for $z\ge0.5$.

\MakeTableee{cr@{\hspace{3pt}}@{$\pm$}@{\hspace{3pt}}lr@{\hspace{3pt}}@{$\pm$}@{\hspace{3pt}}lr@{\hspace{3pt}}@{$\pm$}@{\hspace{3pt}}lr@{\hspace{3pt}}@{$\pm$}@{\hspace{3pt}}lr@{\hspace{3pt}}@{$\pm$}@{\hspace{3pt}}lr@{\hspace{3pt}}@{$\pm$}@{\hspace{3pt}}lr@{\hspace{3pt}}@{$\pm$}@{\hspace{3pt}}l}{12.5cm}{Averaged host ${\rm DM}$}
{\hline
\noalign{\vskip3pt}
z&\multicolumn{2}{c}{${\rm DM}_{\rm IGM}$}&\multicolumn{2}{c}{$\langle{\rm DM}\rangle$}&\multicolumn{2}{c}{${\rm DM}_0$}&\multicolumn{2}{c}{${\rm DM}_3$}&\multicolumn{2}{c}{${\rm DM}_{10}$}&\multicolumn{2}{c}{${\rm DM}_{30}$}&\multicolumn{2}{c}{${\rm DM}_{100}$}\\
\noalign{\vskip3pt}
\hline
\noalign{\vskip3pt}
0.0 &    0 & 0   &  83 & 53  & 147 & 80  &  113 & 54 &  65 & 16 & 37 & 1  & 17 & 0\\
0.2 &  187 & 85  &  93 & 58  & 165 & 84  &  127 & 61 &  75 & 22 & 42 & 2  & 20 & 0\\
0.5 &  493 & 129 & 133 & 72  & 221 & 103 & 156 & 55  & 100 & 18 & 55 & 3  & 23 & 0\\
0.7 &  705 & 157 & 174 & 93  & 271 & 108 & 196 & 60  & 120 & 17 & 60 & 2  & 21 & 0\\
1.0 & 1025 & 197 & 198 & 97  & 296 & 122 & 210 & 63  & 133 & 16 & 74 & 2  & 30 & 0\\
2.0 & 2045 & 293 & 370 & 174 & 460 & 183 & 350 & 121 & 209 & 45 & 80 & 9  & 15 & 1\\
3.0 & 2968 & 345 & 471 & 210 & 513 & 226 & 370 & 174 & 103 & 66 & 12 & 6  &  1 & 0\\
\noalign{\vskip3pt}
\hline
}

\Section{Discussion}
Our results do not support the standard assumptions used when
investigating the plausibility of cosmological tests based on FRBs
with known redshifts. As we have shown the contribution from the host
galaxy to the dispersion measure is in general an increasing function
of the redshift and its standard deviation follows similar path, which
contradicts the hypothesis that both quantities decrease in proportion
to $1/(1+z)$. This is not a fundamental problem, but makes the tests
more difficult, since for a given accuracy one needs larger samples of
FRBs with measured redshifts.

\hglue-4pt Due to the inhomogeneity of the IGM its expected dispersion measure
${\rm DM}_{\rm IGM}(z)$ has some scatter $\sigma_{\rm IGM}(z)$ (compare
Table~1, column~2).  The scatter in the host contribution
$\sigma_h(z)\equiv\sigma_{\rm host}/(1+z)$ (given in the remaining
columns) increases the statistical noise. The combined variance of the
IGM and the host contributions, ($\sigma_{\rm
IGM}^2(z)+\sigma_h^2(z)$), should be compared with the variance of
the IGM part alone ($\sigma_{\rm IGM}^2(z)$). Taking $\sigma_h$ values
from the 3-rd column of Table~1 we see increase of 31\% at $z=0.5$,
and 37\% at $z=3$. Similarly, for the 5-th column (source at 3~kpc
from the host center) we obtain 18\% and 25\% respectively. To preserve
the postulated statistical accuracy of the cosmological tests the
number of FRBs in a sample should increase in proportion. Since this
increase is moderate, the simulations of the accuracy of cosmological
tests based on FRBs under conservative assumptions remain valid.

Our predictions of the host contribution to ${\rm DM}_{\rm obs}$ as shown in
Figs.~3--5 may give some guidance when considering real objects. One
can see that the contribution from galaxies with low stellar mass
$M_*\le10^9~\MS$ is practically unimportant, but it is also unlikely
that they are recognized as hosts at cosmologically interesting
redshifts due to their low luminosity. Looking at the galaxy mass
averaged results one can see that at the projected distance
$R\ge30$~kpc at $z\ge0.5$ the hosts contribution becomes much smaller
than the IGM value or its standard deviation.

Prochaska and Zheng (2019) estimate the dispersion measure of our
Galaxy halo as observed from the Sun position to be
50--80~pc/cm$^3$. Their analysis is based on observations of O VI and
O VII ions which trace the ionized hydrogen.  The stellar mass of the
Milky Way $M_*\approx 6\times 10^{10}~\MS$ (Licquia and Newman 2015)
places it in the highest mass bin considered here.  Our results for a
line-of-sight at 8~kpc from a galaxy belonging to the highest mass bin
at $z=0$ give ${\rm DM}_{\rm host}=64\pm12$~pc/cm$^3$ in an excellent
agreement with the value quoted above.
 
Chittidi \etal (2020) estimate the dispersion measure of the halo of (as
they call it) HG 190608 galaxy, which is the host of FRB 190608, at
$z=0.12$. The galaxy has stellar mass $M_*=2.5\times 10^{10}\MS$ and the
source position is at $R=8$~kpc from the center. Following the methods of
Prochaska and Zheng (2019) the authors estimate the halo contribution to DM
to be in the limits of 30--80~pc/cm$^3$. The galaxy stellar mass places it
in the second highest mass range considered here and close to the boundary
with the highest mass range.  Using our results for the $10^{10}\le M_*\le
3\times10^{10}~\MS$ and interpolating between redshifts we obtain ${\rm
  DM}_{\rm host}/(1+z)=144\pm30$ pc/cm$^3$, about three times higher
estimate. Using the results for the highest mass range we would have $67\pm
12$ (same units) and interpolating in mass and redshift $115\pm 23$. Thus
our prediction of the HG~190608 halo contribution to the observed DM is one
to three times higher as compared to the work of Chittidi \etal (2020)
depending on methodology.

\hglue4pt Bhandari \etal (2020) examine various properties of the four host
galaxies localized with the Australian SKA Pathfinder telescope. These
galaxies have\break $\lg(M_*/\MS)\approx9.4{-}10.4$, and the redshifts
$0.11\le z\le0.48$.  Using our results for $z=0.2$ one would predict their
halo contributions to DM to be (depending on mass) $44\pm15$ to $184\pm78$
[pc/cm$^3$]. Macquart \etal (2020) uses a sample of five FRBs (including four
mentioned above) to estimate the density of baryons in the Universe based on
the observed DM. The host contribution is modeled statistically: it is
assumed that ${\rm DM}_{\rm host}$ has log-normal distribution. The four
parameter model is fitted to the data and gives a weak constraint on hosts:
$\langle DM_{\rm host}\rangle\approx100$ and $50\le{\rm DM}_{\rm host}\le200$
[pc/cm$^3$], which agrees well with our estimates.

The hosts contribution to the observed DM is a technical problem
making cosmological applications of FRBs difficult. With the growing
number of known bursts the correlations of the observed DM with the
number and types of objects near the line-of-sight may give some
limits on ${\rm DM}_{\rm host}$ (compare Prochaska and Zheng 2019). On the
other hand detailed studies of particular hosts, like work performed
by Chittidi \etal (2020) on HG ~90608 should lead to more stringent
relations allowing more accurate estimates of ${\rm DM}_{\rm host}$ based on
other host properties.

\vskip17pt
{\it Note added in proofs:}

After this article had been submitted, the work on the same subject, also
based on {\sf IllustrisTNG} simulation, but using different methods was
published by Zhang \etal (2020).

\Acknow{The {\sf IllustrisTNG} simulations were undertaken with compute time awarded by
the Gauss Centre for Supercomputing (GCS) under GCS Large-Scale Projects
GCS-ILLU and GCS-DWAR on the GCS share of the supercomputer Hazel Hen at the
High Performance Computing Center Stuttgart (HLRS), as well as on the
machines of the Max Planck Computing and Data Facility (MPCDF) in Garching,
Germany.}

\end{document}